\pdfoutput=1

\documentclass[preprint]{elsarticle}

\usepackage[margin=1in]{geometry}    % for 1-inch margins
\usepackage{color, soul}                       % for highlighting.

\usepackage{hyperref}    % for making the references and footnotes clickable
\hypersetup{linktoc=all}

\biboptions{round, semicolon, authoryear}

\usepackage{amsmath}
\usepackage{amsfonts}
\usepackage{amssymb}

\usepackage{tikz}  % for the causal diagram
\usetikzlibrary{arrows, decorations.pathmorphing, backgrounds, positioning, fit, shapes.geometric, arrows.meta}

\usepackage{graphicx}
\newcommand{\indep}{\rotatebox[origin=c]{90}{$\models$}} % to get an "independent" symbol

\begin{document}
\begin{frontmatter}

\title{Causal Inference in Travel Demand Modeling \\ (and the lack thereof)}
\date{June 13, 2016}

\author[tbrathwaite]{Timothy Brathwaite\corref{cor1}}
\ead{timothyb0912@berkeley.edu}

\author[jwalker]{Joan Walker}
\ead{joanwalker@berkeley.edu}

\cortext[cor1]{Corresponding Author}

\address[tbrathwaite]{Department of Civil and Environmental Engineering, University of California at Berkeley\\ 116 McLaughlin Hall, University of California, Berkeley, CA, 94720-1720}
\address[jwalker]{Department of Civil and Environmental Engineering, University of California at Berkeley\\ 111 McLaughlin Hall, University of California, Berkeley, CA, 94720-1720}

\begin{abstract}
This paper is about the general disconnect that we see, both in practice and in literature, between the disciplines of travel demand modeling and causal inference. In this paper, we assert that travel demand modeling should be one of the many fields that focuses on the production of valid causal inferences, and we hypothesize about reasons for the current disconnect between the two bodies of research. Furthermore, we explore the potential benefits of uniting these two disciplines. We consider what travel demand modeling can gain from greater incorporation of techniques and perspectives from the causal inference literatures, and we briefly discuss what the causal inference literature might gain from the work of travel demand modelers. In this paper, we do not attempt to ``solve'' issues related to the drawing of causal inferences from travel demand models. Instead, we hope to spark a larger discussion both within and between the travel demand modeling and causal inference literatures. In particular, we hope to incite discussion about the necessity of drawing causal inferences in travel demand applications and the methods by which one might credibly do so.
\end{abstract}


\end{frontmatter}

\section{What demand modelers have always wanted to do}
\label{sec:travel_demand_goals}
Consider the following three quotations.
\begin{quotation}
``Travel demand models are used to aid in the evaluation of alternative policies. The purpose of the models is to predict the consequences of alternative policies or plans. [...] Predictions made by the model are conditional on the correctness of the behavioral assumptions and, therefore, are no more valid than the behavioral assumptions on which the model is based. A model can duplicate the data perfectly, but may serve no useful purpose for prediction if it represents erroneous behavioral assumptions. For example, consider a policy that will drastically change present conditions. In this case the future may not resemble the present, and simple extrapolation from present data can result in significant errors. However, if the behavioral assumptions of the model are well captured, the model is then valid under radically different conditions.''  ---\citep{ben1973structure}
\end{quotation}

\begin{quotation}
``Indeed, causal models (assuming they are valid) are much more informative than probability models. A joint distribution tells us how probable events are and how probabilities would change with subsequent observations, but a causal model also tells us how these probabilities would change as a result of external interventions---such as those encountered in policy analysis, treatment management, or planning everyday activity. Such changes cannot be deduced from a joint distribution, even if fully specified.'' ---\citep{pearl2009causality}
\end{quotation}

\begin{quotation}
``The goal of many sciences is to understand the mechanisms by which variables came to take on the values they have (that is, to find a generative model), and to predict what the values of those variables would be if the naturally occurring mechanisms were subject to outside manipulations. [...] Finding answers to questions about the mechanisms by which variables come to take on values, or predicting the value of a variable after some other variable has been manipulated, is characteristic of causal inference.''---\citep{spirtes2010introduction}
\end{quotation}

Based on personal communication with many travel demand modelers, i.e. based on anecdote, we believe that the first quotation, by Moshe Ben-Akiva, accurately represents the opinions of most researchers and practitioners within the field of transportation. Moreover, we think it is safe to say that a ``policy that will drastically change present conditions'' can be categorized as an ``external intervention'' or ``outside manipulation.'' If one accepts these two premises, then based on the two quotations by Pearl and Spirtes, it is clear that the implicit goal of travel demand modeling is to make causal inferences (i.e. ``to predict the consequences of alternative policies or plans'')\footnote{As noted by an anonymous referee, ``some might argue that the purpose of demand modeling is to make predictions, as opposed to discover the causal mechanism.'' We believe such distinctions are red herrings. The predominant role of travel demand modelers, especially practitioners, is to predict the effects of particular policies on a future population's travel behavior. As stated by Spirtes (\citeyear{spirtes2010introduction}), ``predicting the value of a variable [i.e. travel behavior] after some other variable [i.e. a policy] has been manipulated is characteristic of causal inference.'' Put succinctly, counterfactual prediction is a causal inference task. Identifying causal mechanisms is also a causal inference task, but identifying causal mechanisms is not always necessary for making counterfactual predictions.}. Moreover, in order to produce such causal inferences, it is clear that travel demand models should be ``causal models.''

In the rest of this paper, we further investigate the relationship between travel demand models and ``causal models'' as seen in other disciplines. Section \ref{sec:brief_primer} provides a brief overview of what causal inference is and why it should be seen as a distinct field from travel demand modeling. In Section \ref{sec:current_state_of_affairs}, we describe the current state of relations between the fields of causal inference and travel demand modeling. There, we pay special attention to the differences between practices in the causal inference literature and practices in travel demand modeling. In Section \ref{sec:why_the_disconnect}, we continue this focus by hypothesizing about why the travel demand modeling literature seems so far removed from the causal inference literature. Finally, although we do not try to ``solve'' the issues of drawing valid causal inferences from travel demand models, we try to bridge the gap between the two literatures in Section \ref{sec:looking_towards_future}. In this section, we emphasize what travel demand modelers can learn from causal inference researchers, we provide an extended example that illustrates the use of the techniques described in this paper, and we conclude with a statement about how travel demand modelers can contribute to the causal inference literature.

\section{A brief primer}
\label{sec:brief_primer}
Despite sharing the same goals (as highlighted in Section \ref{sec:travel_demand_goals}), we believe travel demand modelers are generally uninformed or misinformed\footnote{Note, we do not use the adjectives ``uninformed'' and ``misinformed'' to be disparaging. We mean very literally that travel demand modelers do not seem to widely read the causal inference literature, and because the concepts and findings of that literature are non-trivial and sometimes un-intuitive, travel demand modelers often express sentiments that (1) show a lack of awareness of the technical details and definitions from the causal inference literature or (2) show beliefs that directly contradict findings from the causal inference literature. This second point is supported in the next paragraph.} about key concepts from the field of causal inference.

Here are some recent examples of this point. On April 20th and 21st, 2017, the ``Advancing the Science of Travel Demand Modeling'' National Science Foundation Workshop was held at the University of California, Berkeley. This workshop convened many travel demand modeling scholars and practitioners, young and old, from both within and outside of the United States. As such, the comments made during the workshop represent a wide cross-section of voices within the field. Of special interest was panel discussion \#2: ``How critical is causality? And how can we make clear statements about causality in travel demand models?'' In particular, some direct quotes\footnote{Note that the names of individuals who made each quote have been redacted to respect participant privacy because individuals did not make these statements ``on the record.''} from the discussion after Panel \#2 were:
\begin{itemize}
\item ``What is causality? What is the clear definition of causality?''

\item ``What is causality? What about the context? It's not just $Y$ and $X$.''

\item ``How do we define causality? How much causality is needed in the models to give robust predictions?''

\item ``A model that predicts successfully implies that we are accounting for causality.''

\item ``If we take a certain intervention, will it have the outcome desired by the policy makers? It's not about getting causality right. It's more about what confidence do we have in our projected outcome.''

\end{itemize}
As illustrated by these comments from attendees and the overall tenor of the conversations throughout the workshop, the topic of causality in travel demand modeling is beginning to be widely discussed, but it is still far from being widely and correctly understood\footnote{To be completely explicit, we note that on the topic of making inferences about outcomes under external manipulation or intervention, we generally assume that if the statements of travel demand modelers and causal inference researchers disagree, then the travel demand modeler is incorrect. Of course, we examine the statements and supporting arguments made by both parties, but we have found our assumption to typically hold true. Again, this is not a pejorative remark against travel demand modelers. It is an expected outcome based on the fact that causal inference researchers are trained to focus on this topic, whereas travel demand modelers are typically not.}. Specifically, travel demand modelers seemed most uninformed or misinformed about what causal inference is and how it differs from prevailing practices in travel demand modeling. Below, we briefly address these two questions.

First, for the purposes of this paper, causal inference is defined as the use of data and assumptions to make inferences about outcomes under external manipulation or intervention in a particular context \citep{dawid_2010_beware}. We will begin by introducing some notation. Let $Y_i$ be a discrete dependent variable for an individual $i$. In the field of travel demand, concrete examples of $Y_i$ might be the vector of zeros and ones that represents the travel mode that an individual takes, a count of how many automobiles an individual or household owns, or the time period during which an individual departs from work. Now, let $X_i$ be some explanatory variable for that individual, which is amenable to change via political action. Concrete examples of $X_i$ include the speed of public transit, the cost of driving (as affected by gas taxes), or the prevalence of bicycle lanes between the individual's home and work. Finally, let $Z_i$ be a set of covariates for individual $i$ that also affect the outcome $Y_i$ but are not being subjected to any external change. $Z_i$ might include, for instance, socio-demographics or attributes of a travel mode that are not being subjected to change by the policy in question (e.g. walking time between one's origin and destination).

Using this notation\footnote{Note, our discussion is in terms of a discrete dependent variable because much of travel demand modeling focuses on predicting discrete outcomes. However, if one is interested in a continuous dependent variable, then the quantities described in this paragraph would change as follows. Instead of inferring the probability mass function $P \left( Y_i | \textrm{do} \left( X_i = x \right), Z_i \right)$, we would instead focus on inferring the probability density $f \left( Y_i | \textrm{do} \left( X_i = x \right), Z_i \right)$. Additionally, using $Y_{ij}$ to denote the outcome for individual $i$ if policy $j$ is enacted, we would define the individual causal effect as $Y_{i2} - Y_{i1}$, and we would define the average causal effect as $E \left[ Y_{i2} - Y_{i1} \right]$.}, causal inference focuses on inferring $P \left( Y_i | \textrm{do} \left( X_i = x \right), Z_i \right)$ \citep[Section 3.2.1]{pearl2009statistics}. Here, $x$ is a particular value of $X_i$. The notation $\textrm{do} \left( X_i = x \right)$ explicitly denotes the fact that we are interested in the so-called post-intervention or controlled distribution of the outcome $Y_i$, where we externally set $X_i$ to the value $x$ \citep{pearl_2014_external}. From this post-intervention distribution, numerous quantities of interest may be calculated. For instance, let $X_{i1}$ and $X_{i2}$ be two different values of $X_i$, each corresponding to a different policy: policy 1 and policy 2. Then the individual causal effect of policy 2 versus policy 1 can be defined as $P \left( Y_i | do \left( X_i = X_{i2} \right) \right) - P \left( Y_i | do \left( X_i = X_{i1} \right) \right)$ \citep[Section 8.2.1]{pearl2009statistics}. Other quantities of interest can also be calculated. For instance, the average treatment effect can be defined as the average of the individual causal effects over the population $E \left[ P\left( Y_i | \textrm{do} \left( X_i = X_{i2} \right) \right) - P\left( Y_i | \textrm{do} \left( X_i = X_{i1} \right) \right) \right]$. 

We emphasize here that the post-intervention distributions, i.e. the distributions using the ``do'' operator, contrast the observational distributions of $Y_i$ where individuals choose to have $X_i = x$, e.g. $P \left( Y_i | X_i = x, Z_i \right)$. In general, the two distributions are not equivalent: $P \left( Y_i | \textrm{do} \left( X_i = x \right) \right) \neq P \left( Y_i | X_i = x \right)$. As some readers may already be thinking, this difference is related to the traditionally defined concept of endogeneity. However, as we will discuss two paragraphs from now, this difference in distributions is \textbf{\textit{broader}} than the traditional concept of endogeneity. Now, to clarify what we mean by differences in the post-intervention and observational distributions, imagine the following (fictitious) public health study. Here, $Y_i$ is the number of times an individual rides a bike for recreation in a given month. As an explanatory variable, $X_i$, consider the average number of times in a week that the individual rides his/her bicycle to work. The post-intervention distribution may be observed where participants in a randomized controlled trial are made to ride a bicycle to work 3 or more times a week. This might differ from the observational distribution where perhaps only ``serious'' cyclists rode a bicycle to work 3 or more times a week. Intuitively, one might expect that more recreational bike rides would be observed amongst those who frequently commuted by bicycle without the study intervention as compared to those who were forced to commute by bicycle frequently. In other words, one might expect $P \left( Y_i | X_i = x, Z_i \right) > P \left( Y_i | \textrm{do} \left( X_i = x \right), Z_i \right)$ in this example.

The primary reason for the inequality of the post-intervention and the observational distributions is that individuals choose the values of $X_i$ that they are observed to have. Continuing the example from the last paragraph, individuals choose how often they wish to commute to work by bicycle\footnote{Again, we realize that some readers in the travel demand community may be already thinking that this is simply about endogeneity. Endogeneity, as typically defined, is a \textit{\textbf{subset}} of the concepts used in the causal inference literature when judging the identifiability of $P \left( Y_i | \textrm{do} \left( X_i = x \right) \right)$. We will return to this point in the next paragraph where we will discuss how causal inference is a larger topic than simply dealing with endogeneity as known to travel demand modelers.}. Because individuals choose their observed values of $X_i$, there may be unobserved factors influencing their choice of $X_i$ that also affect their outcome $Y_i$. We discuss this point further in Section \ref{sec:current_state_of_affairs}, but for now, note that in our example, a person with an unobserved aversion to bicycling may still choose not to bicycle a lot for recreation, even if he/she is forced to bicycle to work by his/her doctor. In general, simply looking at the observational distribution $P \left( Y_i | X_i, Z_i \right)$ may lead one to incorrectly overestimate or underestimate the effect of externally setting $X_i = x$ while holding all of the other unobserved variables constant. As put eloquently by statistician A.P. Dawid:
\begin{quotation}
``[I]t is a logically trivial but fundamentally important point that there is no necessary connexion between the different regimes of seeing and doing: a system may very well behave entirely differently when it is kicked than when it is left alone.''---\citep{dawid_2010_beware}
\end{quotation}
As travel demand modelers, we do ourselves a disservice by not paying special attention to this distinction in our modeling efforts. Indeed, our policy problems call for the post-intervention distribution $P \left( Y_i | \textrm{do} \left( X_i = x \right) \right)$, but we typically estimate $P \left( Y_i | X_i = x \right)$. Then, when we apply our models, we erroneously behave as if we have estimated $P \left( Y_i | \textrm{do} \left( X_i = x \right) \right)$. This causal non sequitur leads to misguided statements such as those quoted above where success or confidence in predictions of $P \left( Y_i | X_i = x \right)$ is taken to be the important feature in a causal inference problem, even though the observational distribution may be arbitrarily far from the post-intervention distribution that we truly need.

The discussion above should be helpful for travel demand modelers who are unfamiliar with the field of causal inference and who seek a basic understanding of what the vast and growing body of causal inference studies is about. However, there may be other travel demand modelers who see little added value in the preceding (and following) discussions. Presumably, their thought will be that since the concept of endogeneity and self-selection already exists within travel demand modeling, there is nothing new to be learned. This thought is incorrect. For example, current definitions of endogeneity typically refer to the case where one's explanatory variables are correlated with the error terms in one's model \citep{louviere_2005_recent}. Concretely, imagine that (1) $X_i$ and $T_i$ are two observed, explanatory variables that affect one's outcome $Y_i$, (2) that $X_i$ is currently excluded from one's model while $T_i$ is included, and (3) that $X_i$ and $T_i$ are correlated. Using common definitions, $T_i$ would be labelled endogenous because it is correlated with $X_i$, which is excluded from one's model and therefore part of one's error terms. As such, travel demand scholars who research endogeneity might say that the correct action would be to include $X_i$ in one's model. We will not delve into the details here, but researchers from the causal inference literature would (correctly) point out that the decision of whether or not one should include $X_i$ in one's model depends on one's causal assumptions about how $X_i, T_i,\textrm{ and } Y_i$ are related. In some cases, including $X_i$ can increase bias in one's estimation of $P \left( Y_i | \textrm{do} \left(T_i = t \right) \right)$ instead of reducing it \citep{elwert_2013_graphical, ding_2015_adjust}. For some scenarios where the statistical definitions of endogeneity are insufficient for determining whether a variable should included in one's model, see the literature on M-bias \citep{ding_2015_adjust}, butterfly-bias \citep{ding_2015_adjust}, overcontrol or overadjustment bias \citep{schisterman_2009_overadjustment, elwert_2014_endogenous}, and endogenous selection bias \citep{elwert_2014_endogenous} for more information. 

The basic point that we reiterate and elaborate on further in the remaining sections of this paper is that (1) techniques, approaches, and insights from the causal inference literature are distinct from and broader than those in the current travel demand literature, and (2) that given their common goals, the travel demand literature should both adopt and contribute to methods from the causal inference literature.

\section{The current state of the union}
\label{sec:current_state_of_affairs}
Starting in the 1970's with the so-called Rubin Causal Model \citep{holland1986statistics} and continuing to the present, an impressive amount of scholarly study on causal inference has been performed. This research has largely taken place outside the field of travel demand modeling, within disciplines such as economics, statistics, artificial intelligence/computer science, sociology, and epidemiology. In particular, the causal inference literature has come to focus on a number of discoveries and concepts that are not widely emphasized or utilized within the field of travel demand modeling. The critically important point here is that some of these discoveries in the causal inference literature show that the common viewpoints and practices of travel demand modelers, as exemplified by the Ben-Akiva quotation in Section \ref{sec:travel_demand_goals}, are incorrect or misguided. As a result, the field of travel demand modeling can be improved by incorporating these concerns into its own practice.

Let us give a concrete example to motivate this section. Thus far, much of the causal inference research has focused on the necessary and sufficient conditions for estimating various kinds of causal effects from observational data. Said differently, much causal inference work has focused on specifying the ``requirements for a causal interpretation of an empirical relationship'' \citep{heckman2000causal}. That so much effort has been expended on this topic is instructive. It is now known that having ``valid behavioral assumptions'' that are ``well captured''\footnote{Note that ``well captured'' is taken, here, to mean that one's model is based on and mathematically represents one's behavioral assumptions.} in one's model \textbf{\textit{is not}} sufficient for one to justifiably draw causal inferences from a model estimated from observational data. In the words of \citet{imbens2015causal},
\begin{quotation}
``we cannot simply look at the observed values of [...] outcomes under different treatments [...] and reach valid causal conclusions irrespective of the assignment mechanism. In order to draw valid causal inferences we must consider why some units received one treatment rather than another'' (p.15).
\end{quotation}
We will revisit this notion of treatment assignment later, but for now, the point is that if one estimates a travel model based on ``valid behavioral assumptions'' but fails to consider why the individual decision makers had particular values for the treatment or treatments received (e.g. travel costs and travel times), then one will not be able to make valid causal inferences. To a certain extent, this fact has been acknowledged by academics who work in the field of travel demand modeling \citep{petrin2010control, mabit2010mode, pinjari2011modeling, guevara2015critical}, but such knowledge is not routinely reflected in travel demand research, and it is largely ignored by travel demand modeling practitioners.

Travel models are almost always estimated using observational as opposed to experimental data, and as just noted, there is a discord between the concerns of causal inference researchers and the norms of travel demand modelers. Such a disagreement should spur large changes in how we approach our work as travel demand modelers. In particular, since the predominant purpose of travel demand modeling is to make causal inferences, one might expect travel demand modelers to (as much as possible) have done two things. First, one might have expected modelers to have incorporated the existing causal inference techniques into their own practices. Secondly, one might have expected travel demand modelers to have begun contributing to the general field of causal inference based on their need to make causal inferences in settings that are distinct from the settings typically faced by scholars from other fields. However, despite the two academic disciplines developing roughly simultaneously, no such merger of the causal inference and travel demand modeling worlds has occurred.

To be clear, we recognize that some concepts from the study of causality have made their way into transportation studies. For instance, when trying to determine the effect of the built environment on travel behavior, transportation researchers have long spoken about the ``self-selection'' problem \citep[see for example the review of][]{cao2009examining}. As a specific illustration, consider the impact of transit-oriented-development (TOD) on transit ridership. Here, the issue is that it may not be the presence of TOD that causes higher rates of transit ridership in a given area, but perhaps individuals who prefer to take transit chose to live in TODs. Using terminology from the Rubin Causal Model, one might say that the treatment assignment (TOD or not) mechanism is not random---people choose where to live and therefore choose to be exposed to the treatment. Clearly then, transportation researchers of select topics, such as the land-use and transportation connection or traffic safety, have begun to make use of techniques from the causal inference literature. However, as exemplified by the work done by metropolitan planning organizations, cities, and discrete choice researchers, the general practice of travel demand modeling remains disconnected from the causal inference literature.

Let us provide an example of the disconnect that we are referring to. Treasure Island is between San Francisco and Oakland, California. This island is under the jurisdiction of San Francisco, and a major suite of residential and commercial developments are planned for the island. An important policy objective for San Francisco is that when the initial suite of development is complete, that the majority of travel to, from, and within the island takes place via public transit, walking, and bicycling \citep{treasure2015mobility}. This objective has triggered massive travel demand modeling efforts, both by practitioners and academics. A key piece of these modeling efforts is the creation of travel mode choice models. These models typically take as inputs individual characteristics (e.g. age, gender, family structure, automobile ownership, etc.) and alternative-specific attributes (e.g. travel times and travel costs for a particular individual traveling from a particular origin to a particular destination). As outputs, travel mode choice models return the probability that an individual chooses to complete a trip by a particular travel mode (e.g. car, bus, train, bicycle, walk, taxi etc.). Given a travel mode choice model, as well as a model that can simulate a synthetic population to represent the individuals expected to be living, working, and visiting Treasure Island, one can estimate the expected share of people traveling via each available travel mode. Moreover, such models will be used to study the causal effects on the aggregate travel mode shares, due to the introduction of various types of transportation policies (e.g. transit signal priority, parking restrictions, heavy investments in bicycle infrastructure, etc.).

A major difference between traditional causal inference studies and such travel demand modeling efforts is that there is typically no accounting for the treatment assignment mechanism (i.e. no accounting for confounding\footnote{The term confounding is used in a somewhat technical manner in this paper. Let $C$ denote the set of confounding variables. $C$ may comprise any mix of observed or unobserved variables. Let $Y$ denote the set of outcome variables, and let $D$ denote the set of treatment variables. This paper uses the term confounding to refer to the condition where $C$ has two causal pathways through which it affects $Y$. One is where $C \rightarrow D \rightarrow Y$, i.e. where $C$ affects the value of $D$, that in turn affects the value of $Y$. The second causal pathway is where $C \rightarrow Y$, i.e. where $C$ affects $Y$ through means that do not involve affecting the treatment variables $D$.}) in travel demand modeling work. For instance, the travel mode choice models just described will likely be estimated using disaggregate data collected from household travel surveys. The key parameters being estimated are those that correspond to variables being manipulated by the transportation policies, namely the parameters related to travel times, travel costs, and infrastructure conditions. However, the values of those time, cost, and infrastructure variables were not randomly assigned to the individuals being used for model estimation. Instead, the observed time, cost, and infrastructure values are the result of individuals choosing to live in, work in, and visit particular locations. For instance, since I (Timothy) enjoy commuting by bicycle, I chose to limit my household location search to areas that were within three miles of my workplace. Similar to the TOD example, my choice of bicycling to work is therefore not due solely to having a low bicycle travel time---my bicycle travel time is low because I want to bicycle to work. Put another way, in observational studies such as the kind performed in travel demand modeling, the variables of interest may be endogenous or confounded. Without accounting for this confounding or endogeneity, one has not accounted for the treatment assignment mechanism, and one cannot hope to draw valid causal inferences.

Broadly, we think that a serious problem of the travel demand modeling field is that it ignores findings and methods from the causal inference literature. In particular, travel demand analyses are often not explicit about the causal effects that they are meant to estimate. Moreover, travel demand analyses often lack transparent accounts of how their assumptions and techniques combine to identify the desired causal effects. In the upcoming sections, we will review why we think this gap between the two fields exists, what lessons travel demand modelers can immediately take from the causal inference literature, and where we think travel demand modelers can contribute to the travel demand literature.

\section{Why the disconnect?}
\label{sec:why_the_disconnect}
Given the current state of affairs just described, it may be useful to reflect on why there is a disconnect between travel demand modeling and the study of causality. Below, we state and discuss our (admittedly) subjective views on this topic.

In general, if two academic disciplines address (or appear to address) markedly different problems, then it is quite understandable that those disciplines might not rely on common techniques. For instance, as an extreme example, it is not surprising that creative writing and transportation engineering have very little methodological overlap. These disciplines attempt to answer very different questions. More relevant to this discussion is the fact that travel demand modeling takes place in a setting that is quite different from typical causal inference work. This difference in setting is manifest in terms of the effects or target quantities being studied, how treatments are defined, and the data that is available for use in our studies. We will expound on each of these areas of difference below, but given such differences, it is lamentable---though not surprising---that there is little methodological overlap between causal inference studies and most travel demand modeling efforts. At first glance, travel demand modelers might not think that the causal inference literature will be of much assistance in the sorts of transportation policy questions being addressed.

\subsection{Different Quantities of Interest}
\label{sec:diff_quantities_of_interest}
In terms of the effects or target quantities being studied, questions regarding transportation policy may be ambitious compared to the types of questions typically studied in the causal inference literature. Consider the Treasure Island example once more. The target quantities of interest can be defined as the combined mode shares of public transit, walking, and bicycling under different suites of transportation policies. Given that this is a future development, we observe neither the ``treatment outcome'' nor the reference or control outcome being used as the basis for comparison. Such a setting stands in stark contrast to the typical settings described by prominent researchers of causality. For instance, consider the following three quotes. In ``The State of Applied Econometrics: Causality and Policy Evaluation,'' Athey and Imbens write that
\begin{quotation}
``[w]e focus on the case where the policies of interest had been implemented for at least some units in an available dataset, and the outcome of interest is also observed in that dataset. We do not consider here questions about outcomes that cannot be directly measured in that dataset, such as consumer welfare or worker well-being, and we do not consider questions about policies that have never been implemented. The latter type of question is considered a branch of applied work referred to as ``structural'' analysis; the type of analysis considered in this review is sometimes referred to as ``reduced-form,'' or ``design-based,'' or `causal methods.'"---\citep{athey2016state}
\end{quotation}
Here, Athey and Imbens are explicit about their description of ``causal methods'' not including questions about policies that have never been implemented. Earlier, \citeauthor{imbens2009recent}, in ``Recent Developments in the Econometrics of Program Evaluation'' wrote that
\begin{quotation}
``[t]he central problem studied in this literature is that of evaluating the effect of the exposure of a set of units to a program, or treatment, on some outcome. [...] Moreover, this literature is focused on settings with observations on units exposed, and not exposed, to the treatment, with the evaluation based on comparisons of units exposed and not exposed. As opposed to studies where the causal effect of fundamentally new programs is predicted through direct identification of preferences and production functions.''---\citep{imbens2009recent}
\end{quotation}
And even before this, Nobel Laureate James Heckman wrote that
\begin{quotation}
``[t]he treatment effect literature focuses almost exclusively on policy problem P1 [(evaluating the impact of historical interventions on outcomes)] for the subset of outcomes that is observed. It ignores the problems of forecasting a policy in a new environment [...] or a policy never experienced [...]. Forecasting the effects of new policies is a central task of science and public policy that the treatment effect literature ignores.''---\citep{heckman2005scientific}.
\end{quotation}
The ``treatment effect literature'' that Heckman references is a large subset of the causal inference literature, and these papers are silent about the types of problems that travel demand models are being used for. As a result, travel demand modelers would need to perform a rather substantive search of the causal inference literature to see that some causality work (the so-called structural analysis) is addressing questions that mirror those found in transportation policy analysis.

\subsection{Different Treatments}
\label{sec:diff_treatments}
In much of the standard causal inference literature, the treatment variable in one's analysis is defined as the policy being evaluated. However, in transportation policy analysis, and in the structural analysis segment of the causal inference literature more generally, policies stipulate bundles of treatment variables that are thought to affect one's potential outcomes. For example, when forecasting the effect of a congestion pricing scheme, it is the manipulated automobile travel costs and travel times that will affect one's travel mode choice. Here, the treatment effects of interest are the dose-response relationships between levels of automobile travel costs and travel time, and the probability of an individual choosing to drive.

There are numerous ramifications from redefining treatment variables to be distinct from particular policies. The biggest benefit of this redefinition is noted by Heckman, below.
\begin{quotation}
``This approach models different treatments as consisting of different bundles of characteristics. [...] Different treatments $s$ are characterized by different bundles of the same characteristics that generate all outcomes. This framework provides the basis for solving policy problem P3 [(forecasting the impacts of interventions never historically experienced to new environments)] since new policies (treatments) are generated as different packages of common characteristics, and all policies are put on a common basis''---\citep{heckman2005scientific}.
\end{quotation}

However, despite the increased capabilities brought about by such a redefinition of one's treatment variables, there are at least four drawbacks\footnote{See \citet[Footnote 10]{mokhtarian_2016_quantifying} for a similar discussion of how the structural definition of treatments leads to difficulties in applying standard causal inference techniques in a residential choice setting.}. The first drawback is that to estimate the causal effects of interest, one must now make much stronger assumptions about \textit{how} the treatment variables affect one's outcome variables, as compared to researchers who only study policies that have already been implemented. For instance, instead of simply observing how the construction of transit oriented development changes transit usage rates of residents in an area, one must make assumptions about the mechanisms by which TOD does and does not affect transit usage (e.g. by reducing travel time from the transit station to destinations of interest, but not by making transit usage a more socially acceptable travel mode). In Heckman's words, one must now make assumptions about the ``causes of effects'' instead of simply measuring the ``effects of causes'' \citep{heckman2005scientific}. As a result of discomfort with making such strong assumptions, many scholars who are interested in causal inference do not take the structural analysis approach, and it becomes easy to miss the work of scholars who do focus on forecasting questions that are similar to those seen in transportation.

The second drawback is that while the typical treatment effect literature focuses on categorical treatments (e.g implement one of a finite set of policies), the redefinition described above typically makes use of continuous treatment variables in transportation contexts (e.g. travel times and travel costs). Continuous treatments require one to make even more assumptions in order to arrive at identifiable quantities that can be regarded as treatment effects. In particular, when using the presence of a particular policy as the treatment variable, dummy variables sufficed to describe the treatment effect (e.g. when assuming additive and homogeneous treatment effects). Now, when using continuous treatment variables, one must specify the form of the relationship between the treatment variable and the response (e.g. $x$, $x^2$, $\ln \left( x \right)$, etc.). As before, increasing the number of assumptions that must be made decreases the amount of causal inference literature that is devoted to this sort of transportation-relevant analysis. 

Thirdly, in a ``selection-on-observables'' regime where one believes that he or she has observed all the variables that influence both a person's outcome and his/her observed level of the treatment variables, the redefinition just described may open the analyst up to problems due to the curse of dimensionality. Specifically, many causal inference techniques in ``selection-on-observables'' settings rely on the ``propensity score''---the probability or probability density of the observed treatment level given the observed covariates. When there are multiple treatment variables involved, there are typically multiple propensity scores \citep{imai2004causal}. As the number of treatment variables used to characterize a policy increases, one can encounter a situation with very low numbers of individuals with similar values for all of their propensity scores for the various treatment variables. In such settings, common causal inference techniques such as matching and sub-classification may become difficult to use in practice \citep{imai2004causal}. Travel demand modelers may see such an issue and be initially discouraged, noting that their particular applications suffer from issues that have not even been resolved in the causal inference literature itself.

Finally, as noted earlier, a requirement for drawing causal inferences is that one accounts for the treatment assignment mechanism. Given that the redefinition above typically leads to the creation of multiple treatment variables per policy, travel demand modelers should be concerned about the assignment mechanism for each of the treatment variables for the observations in their sample. Moreover, since confounding due to unobserved variables is usually a serious concern in observational studies with only one treatment variable, it may be reasonable to expect the potential for unobserved confounding to be increased when there are multiple treatment variables. If unobserved confounding exists in one's study, then the prospects for drawing credible causal inferences are grim, partially due to the cross-sectional datasets used in transportation\footnote{For instance, cross-sectional datasets preclude fixed-effects and random-effects estimators that may be used to deal with unobserved heterogeneity.}. 

In the context of travel demand models and cross-sectional data, unobserved confounding shows up as an endogeneity issue. Endogeneity in travel demand models may currently be addressed through a number of techniques such as the use of proxy variables, the ``Berry-Levinson-Pakes'' technique (in particular instances), and instrumental variable techniques \citep{guevara2015critical}. Of these strategies, instrumental variable approaches are the most generally applicable\footnote{As noted by \citet{guevara2015critical}, proxy variable methods are easy to apply but their assumptions are commonly violated. The ``Berry-Levinson-Pakes'' method requires the endogeneity to be present at the level of groups of observations and is not applicable when the endogeneity is present for individual observations \citep{guevara2015critical}.}. Instrumental variable approaches such as control function, latent variable, or multiple indicator solution methods, all rely on researchers being able to find variables that are ``valid instruments.'' That is, one needs variables that are correlated with the endogenous variable but conditionally independent of the outcome, given the endogenous variable(s) \citep{guevara2015critical}. Unfortunately, a travel demand modeler might be dismayed due to the consensus in the causal inference literature that ``[g]ood instruments are hard to find, however, so we'd like to have other tools to deal with unobserved confounders'' \citep{angrist2008mostly}. Even Phillip G. Wright, the inventor of the instrumental variable estimator in econometrics, wrote that ``[s]uch factors, [i.e. valid instruments] I fear, especially in the case of demand conditions, are not easy to find'' \citep{angrist2015mastering}.

\subsubsection{Summary}
In summary, travel demand modelers often hope to draw causal inferences regarding policies that either have not been implemented yet or have not been implemented in the population of interest yet. In such settings, modelers must redefine the ``treatment variables'' in their studies from being particular policies to being sets of characteristics that define policies. This redefinition permits a so-called ``structural analysis'' that is used in a small subset of the causal inference literature. Moreover, this redefinition requires the use of strong assumptions to provide identification of the causal parameters of interest. As a result, the type of causal inference work that most directly pertains to travel demand modelers is not highly visible within the causal inference literature. Additionally, the forays of travel demand modelers into the causal inference literature may not be well received by scholars of the more common ``treatment effect literature'' that do not typically concern themselves with the more speculative studies that are needed in transportation policy analysis. Beyond research visibility and reception, the redefinition of treatment variables may lead to practical difficulties in credibly employing common techniques from the causal inference literature for dealing with confounding/treatment-assignment due to observed or unobserved factors. Such difficulties may be discouraging for travel demand modelers, but they also point to areas where travel demand modeling could contribute to the causal inference literature.

\subsection{Different Datasets}
\label{sec:diff_datasets}
Lastly, as mentioned in the previous subsection, the redefinition of treatment variables, from representing a policy of interest to representing characteristics of policies, may lead to a greater opportunity for an analyst's study to suffer from unobserved confounding. That is, one's treatment variables and one's outcome may both be a function of some unobserved factor(s). Causality researchers, especially economists, have developed a number of techniques for dealing with unobserved confounding, beyond the aforementioned methods. Such techniques include difference-in-difference, fixed effects, and random effect models, to name a few. While this fact may initially seem encouraging to travel demand modelers, these techniques rely on panel data to achieve identification of the causal effects of interest. As already mentioned, travel demand models are typically (though not always) estimated using cross-sectional datasets, thereby precluding the use of many of the existing models for dealing with unobserved confounding.

\subsection{Recapping the rift}
\label{sec:disconnect_summary}
To summarize this section, we have attempted to detail our opinions about why travel demand modeling does not incorporate many of the techniques developed in the causal inference literature. The main reasons that come to mind are that first, not all areas of the causal inference literature are directly applicable to the inferential settings of travel demand modeling. In particular, travel demand modeling often seeks to forecast the effect of a policy that has not been implemented in the target population of interest (e.g. different policies for the future Treasure Island development). Much of the existing causal inference literature ignores this problem in favor of evaluating the effects of policies that have already been implemented in the population of interest. As a result of this difference in questions, a minority subset of the causal inference literature (i.e. the literature on structural analysis) is of greater relevance to travel demand modeling than the more common ``treatment effect literature.''

Secondly, as a result of asking different questions, travel demand modelers will likely need to change their definition of what a treatment variable is. By moving from treatment variables that are equivalent to policies being evaluated, to treatment variables that define characteristics of policies, travel demand modelers are able to draw causal inferences about the effects of policies that have not yet been implemented in the populations of interest. However, in redefining what a treatment variable is, travel demand modelers may face difficulties in applying standard causal inference techniques. For instance, there may be a greater chance of suffering from the curse of dimensionality when applying propensity score techniques. Additionally, there may be a greater need for sensitivity analysis due to modelers making strong assumptions about the nature of the relationship between treatment and outcome variables. And finally, the redefinition of treatment effects may expose modelers to a greater chance of confounding from unobserved factors. Unfortunately, travel demand modeling's ubiquitous cross-sectional data disqualifies many of the tools that have been developed to combat just this type of unobserved confounding.

Put simply, travel demand modeling may not have adopted techniques from the causal inference literature because the relevant techniques are not widely visible, nor are they necessarily straightforward or possible to apply in a transportation setting.

\section{Where we can go from here?}
\label{sec:looking_towards_future}
While the previous section may appear to be a rather somber conclusion about the intermingling ability of the causal inference and travel demand modeling worlds, we are actually quite optimistic that the two fields can actually be mutually beneficial to one another. Presently, we think that there are many practices and perspectives that can be usefully adopted from the various branches of the causal inference literature. We will use Subsection \ref{sec:causal_lessons} to provide an overview of lessons that we think may be most valuable for travel demand modelers. Subsection \ref{sec:final_example} will then illustrate these points on a final, concrete example. It is hoped that this example will be familiar enough to travel demand modelers that they can go forth and begin trying to apply techniques from the causal inference literature in their own work. Finally, we use Subsection \ref{sec:giving_back_to_causality} to conclude by pointing out the potential contributions to the causal inference literature that can come from travel demand researchers.

\subsection{Lessons to learn from the causal inference literatures}
\label{sec:causal_lessons}
This subsection is targeted towards travel demand modelers. Herein, we provide a high-level and subjective overview of what we believe are three key and useful points from the various causal inference literatures. In particular, we make note of topics discussed in the computer science, machine learning, econometrics, statistics, and epidemiology literatures. Where appropriate, we also point out areas that we believe should be of future research interests to the travel demand modeling discipline.

\subsubsection{Lesson 1: Be explicit}
\label{sec:be-explicit}
As expressed by Judea Pearl, ``behind every causal conclusion there must lie some causal assumption that is not testable in observational studies'' \citep[p.99]{pearl2009statistics}. Consequently, travel demand modelers should be explicit about the assumptions they have made in order to draw their conclusions. Such an upfront statement of one's assumptions would facilitate an honest evaluation of the validity of one's claimed causal inferences. In particular, two pieces of information seem key. First, it would be useful for travel demand modelers to explicitly state their assumptions about the causal relationships between the observed explanatory variables, the outcome variables, and the unobserved variables that are thought to affect the outcomes. Secondly, it would be useful for travel demand modelers to explicitly state their identification strategy---i.e., how their dataset and methodology allow them to make use of their causal assumptions to identify the causal relationships of interest \citep{keele2015statistics}. Both of these points will be expounded on below.

In stating one's assumptions about how the observed and unobserved variables of interest are causally related, such statements would ideally be made both graphically and verbally. The figures most frequently used for these graphical displays are directed, acyclic graphs. When used to encode causal assumptions, such graphs are known as ``structural equation models\footnote{Note, structural equation models are often based on linear models \citep{golob_2003_structural}. These parametric assumptions need not be made, and indeed, for the use of encoding causal assumptions we refer to non-parametric structural equation models \citep{bollen_2013_eight} because we are not making any parametric assumptions at this stage of the analysis.}'' in transportation, econometrics, psychology, and sociology \citep{golob_2003_structural, bollen_2013_eight}; ``causal flow diagrams'' or ``system maps'' in systems dynamics \citep{abbas_1994_system, shepherd_2014_review};  ``causal diagrams'' in computer science and systems dynamics \citep{pearl2009causality, abbas_1994_system}; ``influence diagrams'' in statistics \citep{dawid2015statistical}; and ``causal graphs'' or ``path diagrams'' in the social sciences \citep{morgan2015counterfactuals}. These graphs serve multiple purposes. First, they aid one in communicating one's assumptions about a potentially complicated system of relations between various sets of observed and unobserved variables. Additionally, the graphs aid one in determining how and which causal effects are theoretically identifiable given one's assumptions. Once a graph has been shown, a verbal description can follow, explaining any additional causal assumptions, explaining the unobserved variables in greater detail, and/or justifying the exclusion of other variables from the graph.

While the preceding paragraph concerned one's beliefs about how the world works in theory, it is also important to state one's assumption about how the dataset in one's study permits the identification of causal effects. This corresponds to making explicit statements about the details of the dataset being used in one's study and how one's methodology will account for the treatment assignment mechanism of one's observations. As econometricians might say, one should be explicit about where the ``identifying variation'' in one's dataset is coming from and what one's ``identification strategy'' is \citep{angrist2010credibility, keele2015statistics}. Is one saying that all covariates of interest and all confounding variables have been observed? Is one relying on an instrumental variable approach to identification, and if so, what are one's instruments, and how strong are they? How is one dealing with unobserved confounding if any is suspected? These types of questions should be clearly answered in one's study in order to help others judge the validity of one's research.

\subsubsection{Lesson 2: Make fewer assumptions}
\label{sec:fewer-assumptions}
In 1983, Edward Leamer wrote a scathing critique of data analysis practices within economics. Bemoaning the lack of robustness in the conclusions that were drawn from various analyses, Leamer wrote that 
\begin{quotation}
``an inference is not believable if it is fragile, if it can be reversed by minor changes in assumptions. As consumers of research, we correctly reserve judgement on an inference until it stands up to a study of fragility [...]. [...] The professional audience consequently and properly withholds belief until an inference is shown to be adequately insensitive to the choice of assumptions.''---\citep{leamer1983let}
\end{quotation}
Echoing these sentiments, a strong wave of criticism swept the academic world of econometrics and the social sciences more broadly in the 1970's and 1980's \citep{leontief1971theoretical, freedman1985statistics, abbott1988transcending}. The main intellectual thrust of these critiques was that the inferences made by many researchers rested on strong assumptions that could not be credibly defended. As pointed out by econometrician Charles Manski \citep{manski2003partial}, ``the credibility of inference decreases with the strength of the assumptions maintained,'' so based on the dubiously strong assumptions invoked by researchers, scholarly inferences themselves were also deemed untrustworthy.

Within travel demand modeling, where there is nearly ubiquitous appeal to assumptions of utility maximization and Type I extreme value distribution assumptions for unobserved factors, there has been some response to the credibility concerns just mentioned. Discrete choice modelers have relaxed assumptions to allow for taste heterogeneity amongst individuals (mixed logit), substitution patterns across alternatives (nested logit, cross-nested logit, etc.), distributional heterogeneity across alternatives (heteroskedastic logit, mixed logit with alternative specific variances), attribute non-attendance, and more. However, many academic studies, and most travel demand models used in practice, still rely on stringent assumptions about how one's explanatory variables lead to the probability of a given outcome.

In this sense, travel demand modelers may do well to follow the lead of researchers in other disciplines who also conduct model-based causal inference. In disciplines such as econometrics, epidemiology, biostatistics, etc., non-parametric models are beginning to see increased use. These models make substantially weaker assumptions than the assumptions typically made in travel demand models. For example, consider the words of biostatistician Mark van der Laan:
\begin{quotation}
``Why do we need a revolution? Can we not keep doing what we have been doing? Sadly, nearly all data analyses are based on the application of so-called parametric (or other restrictive) statistical models that assume the data-generating distributions have specific forms. Many agree that these statistical models are wrong. That is, everybody knows that linear or logistic regression in parametric statistical models and Cox proportional hazards models are specified incorrectly. [...] However, today statisticians still use these models to draw conclusions in high-dimensional data and then hope these conclusions are not too wrong. It is too easy to state that using methods we know are wrong is an acceptable practice: it is not! [...] That is, instead of assuming misspecified parametric or heavily restrictive semi-parametric statistical models, and viewing the (regression) coefficients in these statistical models as the target parameters of interest, we need to define the statistical estimation problem in terms of non-parametric or semi-parametric statistical models that represent realistic knowledge, and in addition we must define the target parameter as a particular function of the true probability distribution of the data.''---\citep{vanderlaan2011targed}
\end{quotation}

As van der Laan counsels, travel demand modelers should make greater use of non-parametric and semi-parametric models that ``represent realistic knowledge.'' Here, there is likely much room to learn from the practices of modern econometricians who make use of non-parametric models. Likewise, given that the machine learning community builds models of discrete outcomes with minimal assumptions, travel demand modelers can probably benefit from adapting techniques from the machine learning literature. Such a melding of techniques has already begun to occur in other disciplines. For instance, a growing cohort of econometricians and statisticians have begun making use of machine learning techniques for making causal inferences \citep[see for example][]{hill2011bayesian, su2012facilitating, athey2016recursive}, and the fields of epidemiology and biostatistics have begun to do the same \citep[e.g.][]{cruz2006applications, van2010targeted, lee2010improving}. Though machine learning techniques are not widely used within the field of travel demand modeling, we think this can and should change.

\subsubsection{Lesson 3: Validate one's inferences}
\label{sec:validate-inferences}
Undoubtedly, the prospective analyses that are needed in travel demand modeling require a ``structural'' approach to causal inference, where explicit models are used for the probabilities of individual travel choices. However, it would be wise to pay attention to the critiques that have already been levied at the structural approach to causal inference. In particular, it seems prudent to adopt a healthy dose of skepticism towards our travel demand models and subject them to numerous means of validation.

Looking at in-sample means of inferential validation, Leamer writes in a rejoinder to his original critique that ``sensitivity analysis would help'' (\citeyear{leamer1985sensitivity}). We agree. It should be standard practice to subject one's model assumptions to multiple changes (changes in variable specification, radical changes in model form, etc.) in order to assess the robustness of one's results. However, sensitivity analyses by themselves are not enough. As noted by Angrist and Pischke (\citeyear{angrist2010credibility}),
\begin{quotation}
``[a] good structural equation model might tell us something about economic mechanisms as well as causal effects. But if the information about mechanisms is to be worth anything, the structural estimates should line up with those derived under weaker assumptions. [...] We find the empirical results generated by a good research design more compelling than the conclusions derived from a good theory [...].''
\end{quotation}
Such sentiments have been echoed numerous times in the causal inference literature \citep[for e.g.][]{hendry1980econometrics, lalonde1986evaluating}. To ensure that our structural models are producing reasonable inferences, we should also be validating our models using out-of-sample data. Note that this out-of-sample validation does not simply test one's model on more samples from the observational distribution (e.g. such as by hold-out samples or cross-validation). The out-of-sample validation being spoken of here uses samples from a post-intervention distribution where the variables of interest have actually been ``manipulated'' and the samples being used for validation were not part of the original model estimation process.

Such out-of-sample validation can take numerous forms. First, in the case where we are making predictions about some future event (for e.g. travel mode shares on Treasure Island), we should be performing post-evaluations using the actual results that are observed after the event in question (e.g. the actual mode shares after the Treasure Island development is opened). This is reminiscent of the early Bay Area Rapid Transit (BART) studies that were performed by Daniel McFadden \citep{mcfadden1974measurement, mcfadden2000disaggregate}. Before the BART system opened, McFadden predicted BART mode share, and he compared those predictions with the actual mode shares after the system opened. Such comparisons allow one to judge the credibility of a given structural analysis.

Beyond the use of post-evaluation studies, travel demand modelers should take advantage of ``natural experiments'' and highly credible observational studies (e.g. well done regression discontinuity and difference-in-difference designs). For instance, has a transit strike temporarily eliminated the public transit option for travelers? This presents an opportunity to observe whether travelers redistribute themselves according to the patterns predicted by one's travel demand model. Alternatively, is one's city or region considering the implementation of dynamic parking prices? Provided that (1) there is adequate public notice and (2) that prices remain stable long enough for people to reach new equilibrium behaviors, one can observe how people's driving habits change in response to changing driving costs. Do people's real changes match the predictions from one's travel model? Overall, our transportation systems are continually buffeted\footnote{Thanks to Michael Anderson for pointing out the importance of this fact.} by sporadic disturbances that change travel times, travel costs, and various types of physical infrastructure \citep[e.g.][]{marsden_2013_insights}. Such disturbances are invaluable opportunities to observe how well our analyses predict the effects of external changes to these key attributes.

Lastly, one should also strive whenever possible to make use of randomized controlled trials (RCTs). We recognize that there are formidable ethical and logistic challenges to performing RCTs in transportation settings. This is a large part of why RCTs have not been performed more frequently by travel demand modelers. However, as noted by Donald Rubin
\begin{quotation}
``[f]or obtaining causal inferences that are objective, and therefore have the best chance of revealing scientific truths, carefully designed and executed randomized experiments are generally considered to be the gold standard. Observational studies, in contrast, are generally fraught with problems that compromise any claim for objectivity of the resulting causal inferences.'' ---\citep{rubin2008objective}
\end{quotation}
Fortunately, as digital transportation services rise in popularity, the ease with which RCTs can be performed is also increasing. For example, the use of transit smartcards can help transit agencies perform experiments related to transit prices (via electronically distributed discounts) \citep{carrel_2017_san}. Private transportation network companies such as Lyft and Uber already perform large numbers of RCTs on their users, varying attributes such as prices, displayed wait times, etc. \citep{chamandy_2016_experimentation, attwell_2017_engineering}. To the extent that the results of travel demand models built on observational data match the results of these and other RCTs, one can have greater confidence in the inferences from one's model. And critically, if predictions from one's model that was built on observational data does not align with the results of one's RCT, then one should investigate which assumptions need to be modified in order to produce valid inferences.

Importantly, as a result of using post-evaluation, highly credible observational studies of the kind employed in the ``treatment effect'' literature, and RCTs, it often becomes easier to actually implement new transportation policies. The sad, and perhaps justified truth, is that many individuals in the public, many politicians, and even many transportation practitioners do not trust travel model outputs. Based on our experience, travel demand models are often viewed with suspicion. At the same time however, actual data on the result of implemented policies are viewed as having greater credibility. If we are to not just analyze transportation policies but actually be useful in helping good policies get implemented, then evaluation (not just forecasting) must be employed. To this end, consider the following the quote from former New York City Department of Transportation Commissioner Janette Sadik-Khan. Known for her dizzying array of completed projects and change of New York City streets, she wrote that
\begin{quotation}
``like all politics, all transportation is local and intensely personal. A transit project that could speed travel for tens of thousands of people can be halted by objections to the loss of a few parking spaces or by the simple fear that the project won't work. [...] Data showed that interventions that resolved street problems improved safety and had neutral or even positive effects on overall traffic and business. The public discussion slowly graduated from anecdote to analysis. [...] Data change the scope of how we understand the street. They change the question from whether people like or want redesigned roads to whether these redesigns make the street work better.''---\citep{sadik2016streetfight}
\end{quotation}

In sum, post-evaluation, ``treatment effect'' studies, and RCTs are the opposite side of the travel demand modeling coin. All of these actions can help increase model credibility for both analysts and the public, thereby speeding the identification, adoption, and implementation of sound transportation policies.

\subsection{A final example}
\label{sec:final_example}
To end the discussion of what travel demand modelers can learn from the causal inference literature, we will sketch out how one might apply the various lessons from this paper to a travel demand question. In keeping with our stated goal of encouraging discussion and experimentation, as opposed to ``trying to solve issues related to the drawing of causal inferences from travel demand models,'' we do not carry the analysis through. Instead, we merely describe how such an analysis might proceed. This is partially because methodological issues such as those described in Section \ref{sec:why_the_disconnect} remain currently unresolved, and it is beyond the scope of this paper to make such methodological advancements. We emphasize that our example is merely given so travel demand modelers can have a concrete illustration that helps enable them to go forth and begin working out how to use such causal inference techniques in their own research.

The basic problem we will use to illustrate the methods described in this paper is the following. Imagine one is a transportation planner in Berkeley, CA. The policy question of interest is ``if I install a bicycle lane on University Avenue, from the Berkeley Marina to the University of California, Berkeley, how many additional Berkeley residents are expected to commute to work or to school by bicycle?''

Given that this is a question about the effects of an intervention, we are dealing with a causal inference problem. We will first state what we think the steps of analysis might be, and then we will expound on the less familiar steps afterwards. To be clear, we do not think these steps are necessary or sufficient for every causal inference task. However, we think these steps will be useful for and commonly used by many researchers. Now, the steps of analysis might proceed as follows:
\begin{enumerate}
\item
\label{step:re-express-problem}
Re-express the problem in terms of the ``treatment variables'' as described in Section \ref{sec:diff_treatments} instead of the ``treatment policy'' that was initially used to define the problem.

\item
\label{step:draw-diagram}
State one's causal assumptions, both verbally and graphically. This includes drawing the causal diagram that encodes one's belief about how the treatment variables and the other variables of substantive interest in this problem all relate to the outcomes of interest.

\item
\label{step:test-causal-assumptions}
Attempt to falsify the assumptions encoded in the causal diagram by deriving all testable implications from the diagram and testing them on the data at hand. If any of the testable implications are found to be false, then one or more of the assumptions in the causal diagram are false, and we must reformulate our assumptions. If all tests are passed, then the causal assumptions are compatible with the data at hand. Note, however, that we still cannot say the causal assumptions are true.

\item
\label{step:identification-check}
Determine whether or not the desired causal effects are identifiable given one's causal diagram (i.e. given one's causal assumptions about how the world works) and given the type of data one has access to. Note that this involves determining which variables, if any, need to be conditioned on and how the causal effect will be identified.

\item
\label{step:model-building}
Build models for the various quantities that are involved in the expression for one's causal effect. This is the step travel demand modelers are most familiar with and spend the most time on. It includes tasks such as modeling the outcome of interest (e.g. traveler mode choice) as a function of the covariates determined in the previous step.

\item
\label{step:post-evaluation}
Use natural experiments, ``real'' experiments (such as RCTs), or post-evaluation studies to validate one's analysis and determine what, if anything, should be changed about how such analyses are approached in the future.
\end{enumerate}

\subsubsection*{Step \ref{step:re-express-problem}:}
In this example, the treatment policy is the installation of the bicycle lane on University Avenue. However, expressed in this way, the policy is too narrowly defined. Indeed, because there was never a bicycle lane on University Avenue, there is no data on that \textit{\textbf{exact}} policy that can be used to inform our analyses. One cannot, for example, compare the bicycling rate of individuals before and after the installation of the bicycle lane. Instead, we need a variable that can be thought of as representing the mechanism through which all bike lane projects work (not just a lane on University Avenue). For instance, we might (simplistically) hypothesize that installing a bicycle lane on University Avenue affects people solely by changing the percentage of roadways between an individual's home and work that have bicycle lanes on them. Let us define the treatment variable $T_i$ as this percentage\footnote{We understand that in this example, we could have used other treatment variables. For instance we could let the treatment variable be the aggregate quality of the bicycling environment, as measured by the log-sum from a route choice model. We have instead used the treatment variable defined in the text since less background material is required to understand it.}, where ``between'' is some precisely defined region for each individual that is anchored by his/her home and commute destination.

Note that $T_i$ is a function of the policies employed. For each individual, we can define $T_i \left( \textrm{No bike lane} \right) = T_i ^{NBL}$, and this corresponds to the current percentage of roadways with bicycle lanes, since there is currently no bicycle lane on University Avenue. Likewise, we can define $T_i \left( \textrm{Bike lane on University Avenue} \right) = T_i ^{BL}$ as the percentage of roadways with bicycle lanes between individual $i$'s home and destination given that a bike lane on University Avenue is installed. Now, define $Y_i$ as an indicator variable that denotes what travel mode a person uses to commute. The quantity we want to estimate is
$$E \left[ P \left( Y_i = \textrm{bicycle} | \textrm{do} \left( T_i = T_i ^{BL} \right) \right) - P \left( Y_i = \textrm{bicycle} | \textrm{do} \left( T_i = T_i ^{NBL} \right) \right) \right]$$
where the expectation is taken over the entire population of Berkeley.

\subsubsection*{Step \ref{step:draw-diagram}:}
Now, given the well defined problem specified above, we need to draw the causal diagram that depicts our beliefs about how the world works. Guidance on how to construct such diagrams is given in \citep{pearl_1995_causal, greenland_1999_causal, elwert_2014_endogenous, morgan2015counterfactuals}. In order to avoid lengthening this paper, we do not repeat their instructions. The main point, however, is that constructing a causal diagram involves the explicit representation of relationships between the outcome variable $Y_i$, the treatment variable $T_i$, and the miscellaneous other variables that affect $Y_i$---both observed and unobserved.

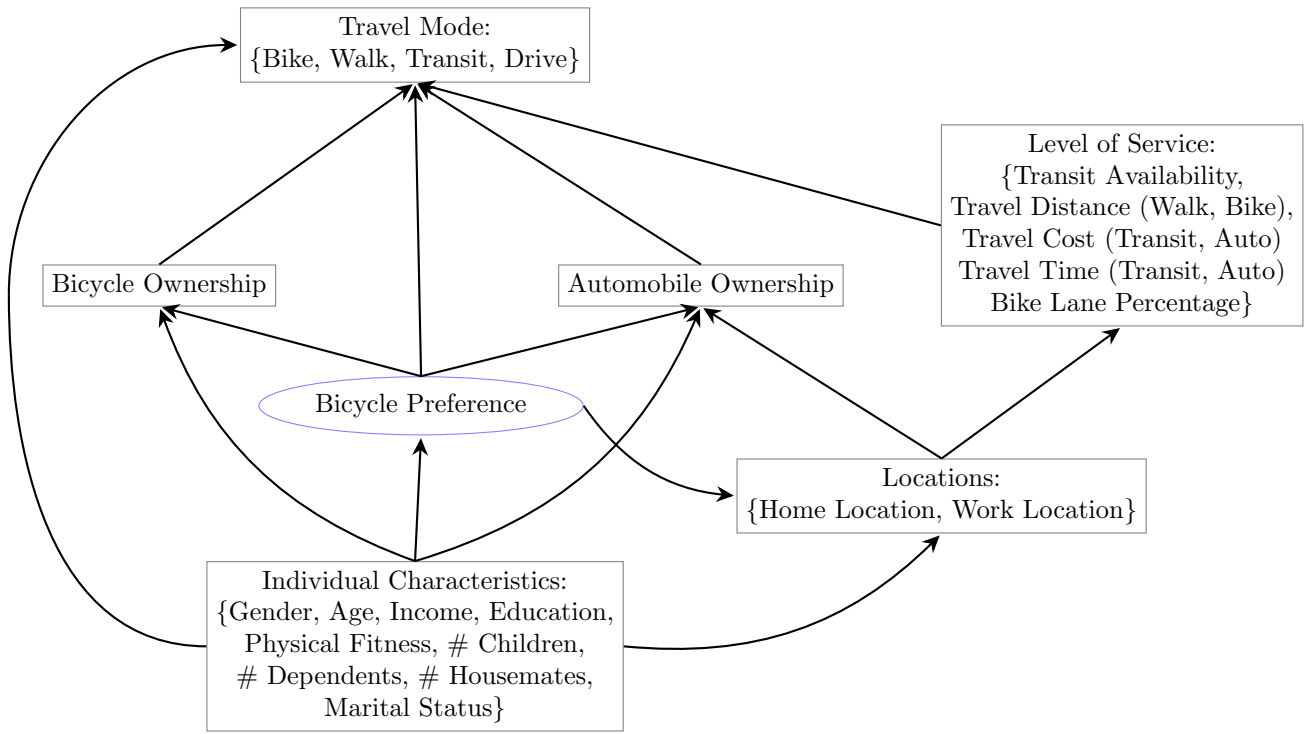
\begin{figure}[h!]
\begin{centering}
\begin{tikzpicture}
  [scale=0.8,
   bend angle=25,
   pre/.style={{Stealth[width=6pt]}-, shorten <=1pt, thick},
   post/.style={ -{Stealth[width=6pt]}, shorten >=1pt, thick},
   %latent/.style={ellipse, draw=blue!50, fill=blue!20, thick},
%   observed/.style={rectangle, draw=black!50, fill=black!20,thick}]
   latent/.style={ellipse, draw=blue!50},
   observed/.style={rectangle, draw=black!50}]

  \node[observed, align=center]   (characteristics)    at ( 1.25, -1)  {Individual Characteristics:\\\{Gender, Age, Income, Education,\\Physical Fitness, \# Children, \\\# Dependents, \# Housemates,\\Marital Status\}};
  \node[latent]         (bike-preference)  at (1.35, 3)  {Bicycle Preference};
  \node[observed, align=center]   (locations)             at ( 10, 1.5)       {Locations:\\\{Home Location, Work Location\}};
  \node[observed]   (bike-ownership)   at ( -3, 5)       {Bicycle Ownership};
  \node[observed]   (auto-ownership)  at ( 6, 5)        {Automobile Ownership};
  \node[observed, align=center]   (level-of-service)  at ( 13, 6)        {Level of Service:\\ \{Transit Availability,\\Travel Distance (Walk, Bike),\\ Travel Cost (Transit, Auto)\\Travel Time (Transit, Auto)\\Bike Lane Percentage\}};
  \node[observed, align=center]   (travel-mode)       at ( 1.25, 9)   {Travel Mode:\\\{Bike, Walk, Transit, Drive\}};
  \node                    (imaginary)          at (-5.5, 5)        {};
  
  \path[->]    (locations.north)              edge [post]    (level-of-service.south)
                                                            edge [post]    (auto-ownership.south)
                   (bike-preference.north)    edge[post]     (auto-ownership.south)
                   									 edge[post]      (bike-ownership.south)
                   									 edge[post]      (travel-mode.south)
                   (bike-preference.east)    edge[post, bend right]      (locations.west)
                   (travel-mode.south)        edge[pre]        (level-of-service.west)
                   									edge[pre]        (auto-ownership.north)
                   									edge[pre]        (bike-ownership.north)
                   (characteristics.north)    edge[post]      (bike-preference.south)
                   								    edge[post, bend right]      (auto-ownership.south)
                   								    edge[post, bend left, bend angle=45]      (bike-ownership.south)
                   (characteristics.east)      edge[post, bend right]      (locations.south)
                   (characteristics.west)    edge[-, thick, out=180, in=270]      (imaginary.south)
                   (imaginary.south)          edge[post, out=90, in=180]   (travel-mode.west);
                   
  \node             (info)     at  (4, -4) {Note:  Squares denote observed variables. Ovals denote unobserved variables.};

\end{tikzpicture}

\caption{Example Causal Diagram for Travel Mode Choice}
\label{fig:bike-causal-diagram}

\end{centering}
\end{figure}

As noted by Morgan and Winship,
\begin{quotation}
``[w]riting down a full graph that represents a consensus position, or a set of graphs that represent alternative positions can be very difficult, especially if the arguments put forward in alternative pieces of research are open to multiple interpretations. Yet little progress on estimating causal effects is possible until such graphs are drawn, or at least some framework consistent with them is brought to bear on the questions of central interest.''---\citep[pg.~33]{morgan2015counterfactuals}
\end{quotation}
One result of this difficulty is that in studies purporting to draw causal inferences, the statement of one's assumptions can be one of the most viciously debated points. Indeed, ``assumptions are self-destructive in their honesty. The more explicit the assumption, the more criticism it invites, for it tends to trigger a richer space of alternative scenarios in which the assumption may fail'' \citep[~pg. 580]{pearl_2014_external}. Here, we do not wish to engage in such debates over our causal assumptions. The example of a causal diagram that we provide in Figure \ref{fig:bike-causal-diagram} is meant to be just that: an example. In a real application, the causal diagram and its embedded assumptions would have to be defended. We simply present such a diagram to give a concrete example of what one might look like and how such a diagram might be used.

When reading Figure \ref{fig:bike-causal-diagram}, note that the variables in the squares are observed, and the variables in ovals are unobserved. In particular, the ``bicycle preference'' node refers to a latent preference for cycling and self-identification as ``a cyclist.'' Additionally, the ``Individual Characteristics'' node and the ``Level of Service'' node denote sets of variables (given in the curly braces in each box). Each of the variables in the curly braces in these two boxes can be thought of as their own node, with the exact same ``parent nodes'' and ``child nodes.'' For instance, both bike lane percentage and transit availability are functions of home and work location, and both bike lane percentage and transit availability influence the travel mode that one chooses to commute by.

As mentioned earlier, the causal diagram encodes one's causal assumptions. For example, Figure \ref{fig:bike-causal-diagram} expresses the assumption that conditional on one's home and work locations,  individual characteristics have no effect on the level-of-service variables. However, not all causal assumptions are displayed by the diagram. One important assumption that is not explicitly shown on the diagram is that the travel mode of a given individual is not affected by the bike lane percentage for other individuals in the population. This assumption of no interference between individuals is known as the Stable Unit Treatment Value Assumption (or SUTVA for short) \citep[p.~10]{imbens2015causal}. SUTVA allows one to estimate causal effects by comparing the probability distributions of the outcomes across groups of individuals with different bike lane percentage levels. Overall, as demonstrated in this paragraph, any causal assumptions of importance that are not encoded in the causal diagram should be explicitly stated in words.

\subsubsection*{Step \ref{step:test-causal-assumptions}:}
Along with a causal diagram come a set of testable implications. Three such types of testable implications are (1) conditional independence tests, (2) tests of functional constraints, and (3) ``over-identification'' tests. First, note that our conditional independence assumptions are encoded in the causal diagram. For instance, given the causal diagram in Figure \ref{fig:bike-causal-diagram}, the following conditional independence\footnote{Note, $a \  \indep \  b \mid c$ means ``$a$ is conditionally independent of $b$ given $c$.''} assumptions are implied:
\begin{itemize}
\item $\textrm{Automobile Ownership } \indep \textrm{  Level of Service} \mid \textrm{Locations}$

\item $\textrm{Bicycle Ownership } \indep \textrm{  Level of Service} \mid \textrm{Locations}$

\item $\textrm{Individual Characteristics } \indep \textrm{  Level of Service} \mid \textrm{Locations}$
\end{itemize}
Each of these assumptions can be tested using one's actual dataset.

Secondly, in models containing latent variables, there may be ``functional constraints'' that can be tested. These constraints are basically statements that certain causal effects (i.e. certain functions) depend only on a particular subset of variables. One can then verify that this is indeed the case by ensuring that the computed causal effect is constant across different values of the variables that are supposed to have no influence on the causal effect. See \citet{tian2002testable} for more information.

Lastly, in a similar fashion, ``over-identification'' can be used when trying to falsify a given causal diagram. ``Over-identification'' refers to the situation where, given a particular causal diagram, there are two or more distinct ways to compute a given causal effect. While this is not the case in the causal diagram of Figure \ref{fig:bike-causal-diagram}, the general idea is that one computes the causal effect by all methods, and then tests for equality of the computed values. See for example \citet{sargan1958estimation, kirby2009using}.

\subsubsection*{Step \ref{step:identification-check}:}
Once one has tried and failed to falsify one's causal diagram, one can perform an identification analysis. This step is now automatic because the necessary procedures have been reduced to algorithms that are implemented in software that is freely available online. See, for example, \citet{breitling2010dagr, textor2016robust, tikka_2017_identifying}. In an observational setting, if the effects one wants to estimate (i.e. $P \left( Y_i = \textrm{bicycle} \mid \textrm{do} \left( T_i = t \right) \right)$) are identifiable, then such software will return an expression for this quantity in terms of observational distributions (i.e. distributions without the ``do'' operator) or a set of variables to be conditioned on in order to estimate the causal effect of interest. By looking at the variables contained in this expression, one will know what variables must be conditioned on, and in looking at the various probability distributions that are returned, one will know what models need to be built and estimated.

For a concrete example, see the diagram given in Figure \ref{fig:bike-causal-diagram} once more. Here, the causal effect is identified, and it is given by the following expression:
\begin{equation}
\label{eq:bike-covariate-adjustment}
P \left( Y_i = \textrm{bicycle} \mid \textrm{do} \left( T_i = t \right) \right) = \int P \left( Y_i = \textrm{bicycle} \mid T_i = t, \textrm{Locations$_i$} \right) P \left( \textrm{Locations$_i$} \mid T_i = t \right) d \left( \textrm{Locations}_i \right)
\end{equation}
From the given expression, one can see that a researcher would need to condition on the home and workplace locations of the various individuals in the dataset, and the researcher would need to build a model for the joint home and workplace location choices of the individual. Here, standard mode choice models are insufficient\footnote{As noted by an anonymous referee, it is not necessarily the case that a standard mode choice model will provide inconsistent estimates of $P \left( Y_i = \textrm{bicycle} \mid \textrm{do} \left( T_i = t \right) \right)$. However, the estimating expressions derived from the ``do-calculus'' operations on causal graphs have already been shown to sufficient for consistently estimating causal effects \citep{galles_1998_axiomatic, huang_2006_pearl}. If analysts wish to use differing expressions, then the analysts should show that their expressions also consistently estimate the desired causal effects or at least meet some other desired criteria.}. Although conventional mode choice models condition on individual characteristics, bicycle ownership, automobile ownership, and the level-of-service variables, controlling for these variables still allows for the possibility that the remaining variation in bike lane percentage is due to the ``confounding'' variable: the individual's latent bicycle preference (through one's home and work location choices). As a result, one cannot treat the observed distribution as being equal to the, desired, post-intervention distribution. We have to directly condition on the home and work locations to sever any ties between the confounding bicycle preference and the bike lane percentage\footnote{Note that if bicycle preference did not directly affect an individual's travel mode choice, then conventional travel demand models would be sufficient for estimating the effect of bicycle lane percentage on mode choice. In that hypothetical scenario, bicycle lane percentage would not be endogenous because bicycle preference would not be part of the error term.}.

\subsubsection*{Steps \ref{step:model-building} and \ref{step:post-evaluation}:}
In the previous subsection, we noted that Equation \ref{eq:bike-covariate-adjustment} called for models of the following probabilities: (1) the probability of bicycling given the bike lane percentage and one's home and work locations, and (2) the joint probability of an individual choosing his/her home and workplace locations, conditional on the bike lane percentage $t$. These models differ from typical travel demand models. The first difference is that the mode choice model, $P \left( Y_i = \textrm{bicycle} \mid T_i = t, \textrm{Locations$_i$} \right)$, conditions on far fewer variables than typical mode choice models. Secondly, the mode choice model directly conditions on the home and workplace locations instead of using proxies such as the level-of-service variables. Thirdly, the full model used to estimate the causal effect combines a joint location choice model with a mode choice model.

Differences from usual travel demand models notwithstanding, at least\footnote{There are definitely multiple sources of analytical difficulty in our example. We are not aiming to be comprehensive. If readers think of their own challenges in this example and wish to know how to address those challenges, we view this as a success for our efforts to spark consideration and discussion of causality in travel demand settings.} two problems will be encountered when trying to construct the needed models. The first issue is the fact that based on subject-matter insight, we know that even in the population, there are few individuals with the same home and workplace location. As a result, one will not have enough individuals to estimate $P \left( Y_i = \textrm{bicycle} \mid T_i = t, \textrm{Locations$_i$} \right)$ after conditioning. Secondly, even if one had multiple individuals with the same home and workplace location, the level-of-service variables such as bicycle lane percentage are a deterministic function of these two locations. There will therefore be no variation in the bike lane percentage after conditioning. Again, the requisite probabilities will not be estimable. Resolving these two issues will be key to solving the causal inference problem given in this example. Such a resolution is beyond the scope of this paper, but we think it is instructive to identify the problem so that other researchers may join us in working on this and similar issues.

As a first step in thinking about how one might identify the causal effects of bike lane percentage on the probability of bicycling to work or school, we offer the following preliminary thoughts. First, while conditioning on variables that influence ``self-selection'' of the bike lane percentage is one way to identify the causal effect of interest, it is not the only way. The identification strategy of conditioning on the variables that lead to bike lane percentage is known as using the ``back-door criterion.'' If we can identify a variable that provides insight into the ``mechanism'' by which bike lane percentage influences an individual's travel mode, then we may be able to use the alternative ``front-door criterion'' \citep{elwert_2013_graphical, knight_2013_causal} to identify the desired causal effect. We will give an example to illustrate this alternative identification strategy. Assume that increasing an individual's bike lane percentage only influences that individual's travel mode by increasing his or her perceived sense of traffic safety for the specific commute trip by bicycle. If we are able to collect measurements of an individual's perceived sense of safety, then using integrated choice and latent variable techniques, we may be able to estimate (1) the effect of bike lane percentage on perceived safety, and (2) the effect of perceived safety on an individual's travel mode. Combining these two estimates with our assumptions, we will be able to estimate the effect of bike lane percentage on an individual's probability of traveling by bicycle. We do not claim that this is the only way to estimate the desired causal effect, or even a correct way to estimate the desired effect (since the assumptions may be incorrect), but we use the discussion as an example of how one might proceed.

Now, once one completes Step \ref{step:model-building}, one will have a model for $P \left( Y_i = \textrm{bicycle} \mid \textrm{do} \left( T_i = t \right) \right)$. This model can then be used to calculate the desired quantity:
$$E \left[ P \left( Y_i = \textrm{bicycle} \mid \textrm{do} \left( T_i = T_i ^{BL} \right) \right) - P \left( Y_i = \textrm{bicycle} \mid \textrm{do} \left( T_i = T_i ^{NBL} \right) \right) \right]$$ The end result will be a causally valid inference about the effect of the University Avenue bike lane on citywide demand for bicycle commuting.

For Step \ref{step:post-evaluation}, one should use data from actual bicycle lane interventions to corroborate the model that one is making inferences from. Note that evaluating real outcomes to see whether or not they match one's predictions has been a part of travel demand analysis from the beginning (see the discussion in Section \ref{sec:validate-inferences} about the early BART studies). While such evaluations may not be performed regularly by travel demand modelers any more, the knowledge of how to do so exists. See for example, Section \ref{sec:validate-inferences}. In the context of our example, natural experiments might take the form of measuring bicycle usage before and after the construction of bicycle lanes that are built in an ``unexpected'' manner. For instance, looking before and after the stealthy construction of bicycle lanes in New York City for public trials. We can then compare our predictions of the change in bicycle mode share before and after the installation of these lanes. Alternatively, a RCT might be analyzed whereby low-income residents applying for housing assistance are randomly placed in housing and the individuals have differing values of $T_i$. Bicycle commuting rates can then be studied using the different individuals in the program. Finally, if the bicycle lane is actually constructed on University Avenue, one should perform a post-evaluation study whereby the bicycling rates amongst residents are measured before and after the construction of the bicycle lane\footnote{Of course, care should be taken to measure bicycling rates amongst those who lived and worked in the area for a sufficiently long time before and after the construction of the bike lane. This would be done to remove any influence of individuals who may have moved their home or work location to the area after they knew the bike lane existed or was going to be built.}. Such studies will confirm whether one's model is actually performing well.

\subsection{Conclusion}
\label{sec:giving_back_to_causality}
So far, we have discussed what causal inference is, the overlapping goals of causal inference and travel demand modelling, and some reasons why we think a gulf exists between these two disciplines. Moreover, we have extracted some lessons from the causal inference literature that we think can be of use for travel demand modelers, and we have tried to show how these lessons might be used in a concrete, travel demand setting. To conclude, we now turn to the prospect of the causal inference literature being enriched by the work of travel demand modelers, and we end with a distinctly hopeful outlook.

Travel demand modeling, in its modern incarnation, grew out of the application of econometrics to the study of human travel patterns. The problems faced in modeling human travel choices are difficult, and as a result, travel demand modeling applications provided the impetus for many of the most advanced discrete choice modeling techniques to date. Concurrently, the broader field of econometrics has moved on to embrace the challenge of determining causal effects from observational and experimental data \citep{angrist2010credibility}, and we think it is only natural that travel demand modeling ``catch-up'' to its progenitor. As noted in Sections \ref{sec:why_the_disconnect} and \ref{sec:causal_lessons}, there are a number of challenges to be faced in bringing causal inference techniques and perspectives to bear on travel demand modeling applications. However, methodological challenges have always provided the most fertile ground for progress. Accordingly, we highlight three causal inference topics that we think will be particularly fruitful grounds for research and application by travel demand modelers.

First, causal inference researchers in computer science and epidemiology have been producing a small but growing literature on the topic of ``causal transportability'' and ``meta-synthesis'' \citep{hernan_2011_compound, petersen_2011_compound, pearl_2011_transportability, bareinboim_2012_transportability, pearl_2012_calculus, lee_2013_m, bareinboim_2013_transportability, singleton_2014_motivating, bareinboim_2014_transportability}. Put simply, the study of causal transportability seeks to determine the conditions and procedures with which it is possible to \textit{transport} (i.e. generalize) causal inferences learned in one setting to another \citep{pearl_2011_transportability, pearl_2014_external}. Similarly, ``meta-synthesis'' \citep{pearl_2012_calculus, lee_2013_m, bareinboim_2013_transportability, bareinboim_2014_transportability} is concerned with the formalization of procedures for combining inferences from multiple studies into one aggregated measure for a target population that need not have been involved in any of the studies being combined. Thus far, papers about causal transportability have mainly focused on the abstract mathematical conditions that permit the transport of causal inferences. There has been a comparative lack of research applying the knowledge obtained thus far to real applications\footnote{Notable exceptions include work such as \citet{singleton_2014_motivating}.}. There is likely much to be gained from applying the abstract mathematics of causal transportability research to real problems and from attempting to integrate transportability notions with domain specific modeling techniques and traditions. Already, travel demand modelers have much experience with two specific transportability processes: (1) transferring model results from one time and place to another \citep{agyemang_1997_spatial, fox_2015_temporal}, and (2) generalizing insights from stated preference (SP) experiments to revealed preference (RP) studies. In the latter case, travel demand studies have used a number of techniques to facilitate the desired transport of model results. For example, these techniques include methods such as joint RP-SP estimation techniques \citep{brownstone_2000_joint, feit_2010_reality}, incentive-aligned SP experiments to increase the similarity between the SP study and the RP environment where the results will be used \citep{ding_2005_incentive, ding_2007_incentive, moser_2010_using, chung_2017_willingness}, and certainty calibration \citep{beck_2016_can}. From both model transferability and RP/SP studies, travel demand modelers may have much accumulated wisdom to offer the causal transportability literature. Conversely, travel demand modelers may also have much to gain by incorporating the existing causal transportability techniques, especially when trying to determine whether or not the desired inferential transport can actually be performed.

Secondly, feedback processes and change over time have been ignored in much of the causal inference research performed thus far\footnote{Thanks to two anonymous referees for raising this point.}. Indeed, much of the causal inference work performed thus far takes the directed \textit{acyclic} graph as its starting point (or equivalently, the \textit{recursive} structural equation model used in much of the social sciences). Such work explicitly excludes systems of relationships where variable 1 both causes and is affected by variable 2 (possibly offset by a time lag). A transportation example of such a feedback process is where an individual's attitudes towards bicycling affect the individual's choice of bicycling or not, but the individual's experiences while bicycling will then affect his/her future attitudes towards bicycling. While possibly a rare occurrence in other disciplines, we expect that such feedback processes are common in travel demand settings. As another example, consider the effect of increased driving costs. At the outset, the increase in costs is expected to damp demand by some initial amount. However, the initial decrease in the number of drivers may alleviate congestion, thereby causing an increase in traffic speeds and leading some potential drivers to begin driving due to the faster speeds. The overall decrease in the number of drivers will therefore be less than the initial decrease due to the increased prices. This overall decrease in the number of drivers may be over-predicted if the feedback mechanism is ignored. To give credit where it is due, a limited amount of causal inference work has tried to account for such feedback processes. This work includes the use of chain graphs \citep{lauritzen_2002_chain}, directed \textit{cyclic} graphs \citep{schmidt_2009_modeling}, the ``settable systems'' framework developed by econometricians Halbert White and Karim Chalak (\citeyear{white_2009_settable}), and dynamic causal networks \citep{blondel_2017_identifiability}. However, as with the aforementioned causal transportability studies, these techniques have seldom been used in real applications. Here, travel demand modelers would essentially forge the link between systems dynamics researchers who commonly use causal diagrams to portray systems with feedback processes over time (for example \citet{abbas_1994_system} and \citet{shepherd_2014_review}) and causal inference researchers who use causal diagrams to explicitly identify and compute causal effects.

Finally, travel demand researchers face severe challenges when trying to make robust quantitative claims. The fragility of travel demand modeling results often comes from ambiguity over how to choose the variables to be conditioned on when trying to estimate the probability of interest, how to specify the ``systematic utility'' equations\footnote{We realize that not all travel demand models have systematic utilities. However, given that most travel demand models are based on random utility maximization, the comments in this paragraph are valid for most travel demand analyses.}, how the preferences underlying the systematic utilities might change over time, how to specify the probability function that links the systematic utilities with the probability of the observed choice, and even ambiguity in the causal diagram upon which the entire analysis should be built (for e.g. see the literature on observationally equivalent causal graphs). As was recently noted by Dagsvik:
\begin{quotation}
``A well-known problem in quantitative economic analysis is that economic theory provides limited guidance for the specification of functional forms of quantitative structural economic models. An unfortunate consequence is that it becomes difficult to discriminate between econometric model formulations based on the same theoretical framework which fit the data reasonably well but result in different counterfactual predictions. Given this state of affairs, the analyst is forced to choose between model specifications without adequate theoretical or empirical support.''--\citep{dagsvik_2017_invariance}
\end{quotation}
Overall, we believe that this ambiguity in travel demand models will remain for the foreseeable future. So although the type of sensitivity analysis advocated in Section \ref{sec:validate-inferences} will help uncover how much uncertainty in one's estimates there actually is, one will likely always need good methods of reporting this uncertainty. Here, we have seen few \textit{structural} causal inference studies that managed to present the uncertainty inherent in their analysis in an easy to understand and useful manner. In our opinion, the inherent ambiguity in travel demand models gives transportation researchers an opportunity to take the lead on discovering parsimonious and meaningful ways of representing the uncertainty in one's analysis.

In conclusion, if the past is any indication of the future, then based on the topics above, we believe that travel demand modelers will once again be able to advance the arsenal of quantitative techniques, this time in the causal inference arena as opposed to simply in the arena of discrete choice. By generalizing and adapting causal inference techniques for travel demand applications, travel modelers can simultaneously contribute to the field of causal inference and fulfill their original purpose of \textit{validly} ``predict[ing] the consequences of alternative policies or plans [...] under radically different conditions'' than those present at the time of one's analysis.

\section*{Acknowledgements}
We thank Elias Barenboim, Kenneth Train, Eric Miller, Mohamed Amine Bouzaghrane, and Hassan Obeid for comments on earlier drafts of this paper, and we thank Michael Anderson for providing us with early introductions to the topic of causal inference. We are also very grateful to our anonymous reviewers for their thoughtful and detailed comments on an earlier draft of this manuscript. Nevertheless, any viewpoints (and, of course, any errors) expressed in this manuscript are entirely our own.

\newpage
\section*{\refname}
\bibliography{causal_transportation}

\end{document}